\documentclass [12pt]{article}
\usepackage {graphicx}
\usepackage {longtable}

\begin{document}
\setlength{\baselineskip}{16pt}
\title{Induced matter: Curved N-manifolds encapsulated in Riemann-flat
N+1 dimensional space}
\author { {Harry I. Ringermacher} \\ {General Electric Global Research Center}\\
{Schenectady, NY 12309}\\ \\ 
{Lawrence R. Mead} \footnote {Communications to lawrence.mead@usm.edu}\\
{Dept. of Physics and Astronomy} \\{University of Southern Mississippi}\\
{Hattiesburg MS. 39406} } 
\date{\today}
\maketitle
\begin{abstract}
Liko and Wesson have recently introduced a new 5-dimensional induced matter 
solution of the Einstein equations, a negative curvature Robertson-Walker space
embedded in a Riemann-flat 5-dimensional manifold. We show
that this solution is a special case of a more general theorem prescribing
the structure of certain N+1 dimensional Riemann-flat spaces which are all
solutions of the Einstein equations. These solutions encapsulate N-dimensional
curved manifolds. Such spaces are said to \lq\lq induce matter" in the 
sub-manifolds by virtue of their geometric structure alone. We prove that the 
N-manifold can be {\em any} maximally symmetric space.
\end{abstract}
\newpage
\setlength{\baselineskip}{16pt}

The concept of \lq\lq induced matter", was originally introduced by 
Wesson~\cite{1,2}.
While investigating 5D Kaluza-Klein theory, he recognized that a curved 
4-space could be embedded in a Ricci-flat ($R_{AB}=0$; 
$A,B,...\in \{0,1,2,3,4\}$) 5-space. This is a reflection of the 
Campbell-Magaard theorem~\cite{3} which, applied to 5D, states that it is 
always possible to embed a curved 4D manifold in a 5D Ricci-flat space. 
Seahra and Wesson~\cite{4} provide an overview and rigorous proof of the 
Campbell-Magaard theorem with applications to higher dimensions. Wesson takes 
\lq\lq induced matter" to mean that the left-hand geometric side 
extra terms of the flat 5D Ricci-tensor provide the source terms in the 
4D curved Ricci-tensor of the embedded space. A \lq\lq weak" version of this 
concept utilizing an embedding of the Friedmann-Roberston-Walker(FRW) 4-space 
in a Minkowski 
5-space has been used to visualize the big bang sectionally~\cite{5}. 
Here, the 5-space is Riemann-flat ($R_{ABCD}=0$) since it is Minkowski. 
There is no physics 
in the 4D subspace, except with reference to the original FRW coordinates.
This simply provides a Euclidean embedding diagram. 

More recently, Liko and Wesson have introduced a new 5D, Riemann-flat 
solution~\cite{6} which they found could \lq\lq encapsulate" a 4D curved  FRW 
space. We use the term \lq\lq encapsulate" as distinct from embed since in 
this 5-space, the coordinates are not Minkowski. The 4D subspace is itself 
curved in the same 5D coordinates.
It is true that a flattening transformation can be found to a 5D Minkowski
space. However, this would simply produce another embedding diagram. The 
physics seems to lie in the encapsulating 5D metric. We shall use the term 
"induced matter" to include a Riemann-flat 5D manifold encapsulating a 
curved 4D subspace. The Liko-Wesson induced matter solution goes on to 
describe an apparently inflationary universe as a negative curvature FRW space embedded in a special 5D universe. The RW space undergoes accelerated 
expansion subject to a repulsive \lq\lq dark energy" ($P=-\rho$). 
We will show in this paper that the Liko-Wesson solution is a special case of 
a more general class of maximally 
symmetric sub-manifolds embedded in Riemann-flat space. A detailed discussion 
of maximally symmetric sub-manifolds based on Poincar\`e metrics and their 
consequences can be found in ref.[7]. For convenience, we repeat some critical definitions and calculations.

Consider the Riemann manifold defined by
\begin{equation}
dS^2=\tilde g_{ij}dx^i dx^j.
\end{equation}
This space is said to be {\em maximally symmetric} if and only if it has
constant {\em sectional curvature} $\kappa=\kappa(i,j)$, for any $1\leq
i\not = j\leq N.$ In the plane spanned by the basis vectors $(\hat e_i,\,
\hat e_j)$ the sectional curvature is defined by,
\begin{equation}
\kappa(i,j)=\tilde g^{ii}R^j_{\,iji} \qquad \hbox{($i,j$ not summed)}.
\end{equation}
For a maximally symmetric space $R^j_{\,iji}=\kappa \tilde g_{ii},$ 
$j\not = i$. For such a space,
\begin{equation}
R_{ii}=-\kappa(N-1)\tilde g_{ii}.
\end{equation}
\noindent{\bf Theorem}: Let $\tilde g_{ij}$ represent a maximally symmetric
space of sectional curvature $\kappa$. {\em The metric}
\begin{equation}
dS^2=d\tau^2-D\tau^2\tilde g_{ij}dx^i dx^j, \quad \hbox{$i,j\dots,\in \{
1,2,\dots,N \}$ },
\end{equation}
{\em is Riemann-flat whenever} $D=-\kappa$.
\\ \\
\noindent{\bf Proof}: Consider the metric 
\begin{equation}
dS^2=d\tau^2-f(\tau)^2\tilde g_{ij}dx^i dx^j, 
\end{equation}
where $\tilde g_{ij}$ denotes a maximally symmetric space.
We compute the independent components of the curvature tensor (the overprime
denotes differentiation in $\tau$):
\begin{eqnarray}
R^i_{\ 0j0} &=& -{f''\over f} \delta^i_{\ j} \nonumber \\
R^0_{\ i0j} &=& - f f'' \tilde g_{ij} \nonumber \\
R^k_{\ ikj} &=&\tilde R_{ij} - (N-1)f'^{\,2} \tilde g_{ij} \\
            &=&-\kappa (N-1)\tilde g_{ij}-(N-1)f'^{\,2}\tilde g_{ij} \nonumber \\
            &=&-(N-1)(f'^{\,2}+\kappa)\tilde g_{ij},\nonumber
\end{eqnarray}
where we have made use of the result Eq.(3). It is evident from Eq.(6) that
the space will be Riemann-flat if and only if $f''=0$ and $f'^{\,2}+\kappa
=0$. Let $f(\tau)=\sqrt{D}\,\tau.$ Then $f''=0$ and $f'^{\,2}-D=0$. It
follows that $D=-\kappa$ and the proof is complete.

Liko and Wesson~\cite{6} introduce the line element (with overall sign of 
$dS^2$ reversed from ours),
\begin{equation}
dS^2=d\tau^2-{\tau^2\over L^2}\bigg [dt^2-L^2\sinh^2({t\over L})d\sigma_3^2
\bigg ],
\end{equation}
where,
\begin{equation}
d\sigma_3^2=\bigg (1+{kr^2\over 4} \bigg )^{-2}(dr^2+r^2d\theta^2+
r^2 \sin^2\theta d\phi^2), \nonumber
\end{equation}
is the Robertson-Walker 3-space with $k=-1$. 

Define coordinates $x^A=\{ \tau,r,\theta,\phi,t\},\ A\in\{0,1,2,3,4\}.$

We can then identify in Eq.(4),
\begin{equation}
\tilde g_{ii}=\{-f,-fr^2,-fr^2\sin^2\theta,1\},\qquad 
f=L^2\sinh^2(t/L)\biggl (1+{kr^2\over 4}\biggr )^{-2}, \nonumber
\end{equation}
and $D=1/L^2$.
Eq.(7) is thus of the form (4) and will satisfy the theorem  
provided that the sectional curvature of the 4-space is $\kappa=-1/L^2$.
Direct evaluation of the sectional curvature for two typical cases (by
symmetry, the remaining cases are identical) results in,
\begin{eqnarray}
\kappa(1,2) & = & \tilde g^{11}R^2_{\,121}  =  -{1\over L^2} \nonumber\\
\kappa(4,2) & = & \tilde g^{44}R^2_{\,424}  =  -{1\over L^2} \nonumber
\end{eqnarray}
which show that the conditions (6) are met. That is, the 4-space
has constant sectional curvature which then results in a Riemann-flat
5-space. We have thus shown that the new metric solution Eq.(7) introduced
by Liko and Wesson is a special case of our more general theorem Eq.(4)
which allows the N-space to be any maximally symmetric manifold.

\end{document}